\documentclass[journal=jacsat,manuscript=article]{achemso}

\usepackage[version=3]{mhchem} 
\usepackage{multirow}
\usepackage{graphicx}
\usepackage{dcolumn}
\usepackage{subfigure}
\usepackage{wrapfig}
\usepackage{caption}
\usepackage{subcaption}
\author{M. A. Keskiner}
\affiliation[Bilkent University]
{Department of Physics, Bilkent University, Ankara, 06800, TURKEY}
\email{akif.keskiner@bilkent.edu.tr}
\author{Pouyan Ghaemi}
\affiliation{Physics Department, City College of the City University of New York, NY 10031, U.S.A.}
\alsoaffiliation{Physics Program, Graduate Center of City University of New York, NY 10031, U.S.A.}
\email{pghaemi@ccny.cuny.edu}
\author{M. \"{O}. Oktel}
\affiliation[Bilkent University]
{Department of Physics, Bilkent University, Ankara, 06800, TURKEY}
\email{oktel@fen.bilkent.edu.tr}
\author{Onur Erten}
\affiliation[Arizona State University]
{Department of Physics, Arizona State University, Tempe, AZ 85287, USA}
\email{onur.erten@asu.edu}

\title{Theory of moir\'e magnetism and multi-domain spin textures in twisted Mott insulator - semimetal heterobilayers}


\begin{document}



\begin{abstract}
Motivated by the recent experimental developments in van der Waals heterostructures, we investigate the emergent magnetism in Mott insulator - semimetal  moir\'e superlattices by deriving effective spin models and exploring their phase diagram by Monte Carlo simulations. Our analysis indicates that the stacking-dependent interlayer Kondo interaction can give rise to different types of magnetic order, forming domains within the moir\'e unit cell. In particular, we find that the AB (AA) stacking regions tend to order (anti)ferromagnetically for an extended range of parameters. The remaining parts of the moir\'e unit cell form ferromagnetic chains that are coupled antiferromagnetically. We show that the decay length of the Kondo interaction can control the extent of these phases. Our results highlight the importance of stacking-dependent interlayer exchange and the rich magnetic spin textures that can be obtained in van der Waals heterostructures.
\end{abstract}

Kondo model is a quintessential strongly-correlated model that describes the interaction between localized magnetic moments and weakly-interacting conduction electrons via an antiferromagnetic spin-spin interaction\cite{Stewart_RMP1984, Coleman_book, Wirth_NatRevMat2016}. Commonly utilized for lanthanide and actinide intermetallic compounds, the Kondo model can possess diverse phenomena such as heavy fermion formation, magnetic order, and unconventional superconductivity \cite{Stewart_RMP1984}. Unlike other interacting models, such as the Hubbard model, the Kondo model exhibits a clear separation between the localized and the itinerant degrees of freedom, which allows for controlled calculations in certain limits\cite{Coleman_book}. In the simplest form, the Kondo model involves only local interactions since both the localized moments and the conduction electrons originate from the $f$ and $d$ orbitals of the same lanthanite/actinide ions in most intermetallics. 

\begin{figure}
\centering
   \includegraphics[width=1\textwidth]{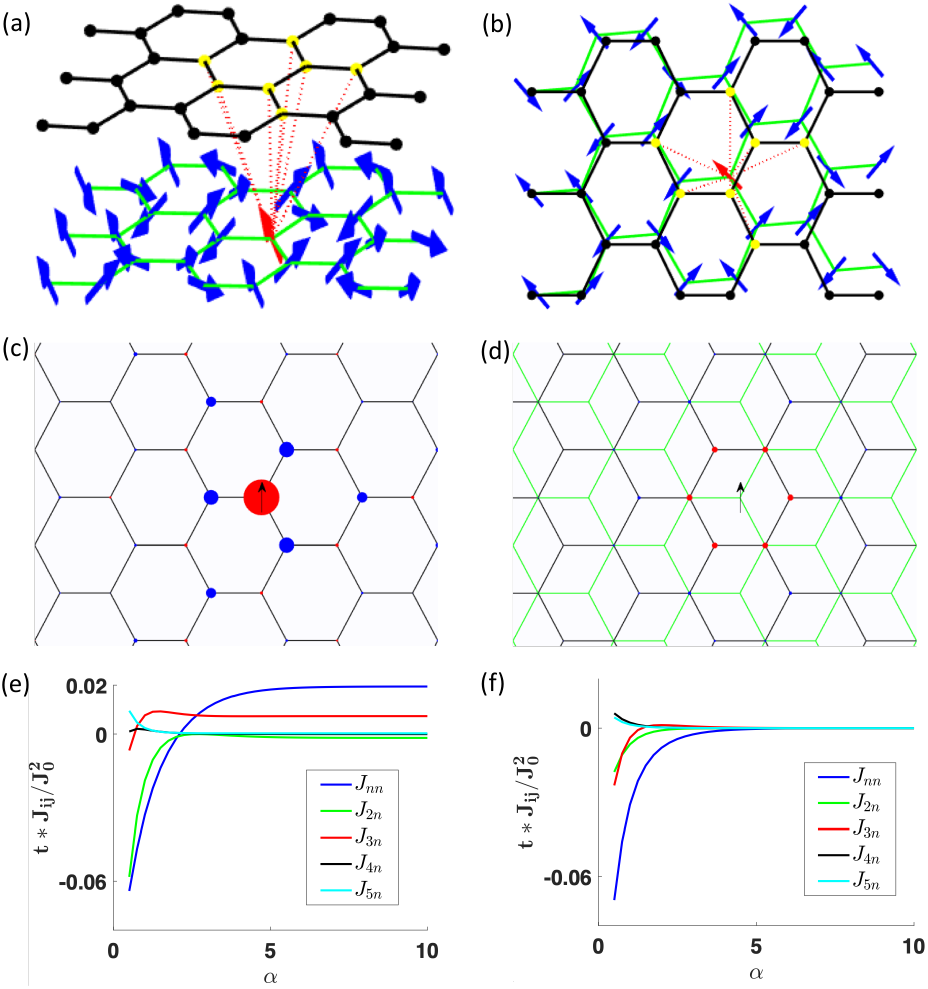}
    \caption{(a) Schematic representation of our model with conduction electrons on the upper layer and localized magnetic impurities on the lower layer, twisted relative to the upper layer. The magnetic impurity, indicated by the red arrow, interacts only with the conduction electrons at the sites shown in yellow in the conduction layer. (b) Top view of our schematic. (c), (d) Induced spin polarization for AA and AB stacking respectively for the inverse decay length $\alpha =3$. The impurity spin is marked with black arrow and the area of the circles on each site is proportional to the spin polarization, where excess spin-$\uparrow$ (spin-$\downarrow$) density is blue(red). (e), (f) The RKKY coupling, $J$ as a function of $\alpha$ for AA and AB stacking up to fifth nearest neighbor.}
    \label{Fig1}
\end{figure}

Recently, an alternative route to realize the Kondo model has emerged in van der Waals (vdW) heterostructures. These synthetic Kondo lattices consist of a two-dimensional (2D) Mott insulator layer and a metallic or semimetallic layer as depicted in Fig.~\ref{Fig1}(a,b). This route separates the Kondo model's two necessary degrees of freedom into different layers, in contrast to the intermetallics. For instance, a synthetic Kondo lattice has been realized in 1T-TaS$_2$/2H-TaS$_2$ bilayers where 1T-TaS$_2$ is a Mott insulator, and 2H-TaS$_2$ is a metal\cite{Ayani_Small2023}. STM experiments show a Kondo resonance in the tunneling spectrum appearing at around 27K\cite{Ayani_Small2023}. Similarly, synthetic Kondo lattices may be constructed in 2D magnet-metal/semimetal bilayers. As an example, a Kondo model with Kitaev intralayer interactions has been proposed for graphene/$\alpha$-RuCl$_3$ bilayers\cite{Jin_PRB2021}. Recent experimental and theoretical studies show that the interaction between graphene and $\alpha$-RuCl$_3$ can have a significant impact on the electronic and magnetic properties of both layers\cite{mashhadi2019,zhou2019,biswas2019,Rizzo_NanoLett2020,gerber2020,leeb_PRL2021,balgley2022,shi2023magnetic}.

Compared to intermetallics, 2D vdW materials offer a broader tunability through gating, electric field, pressure, and strain\cite{Liu_NatRevMat2016}. Their properties can be further controlled by twisting in moir\'e superlattices\cite{He_ACSNano2021}. A lattice mismatch in heterobilayers can also form moir\'e superlattices, even without twisting. For instance, the in-plane lattice constant of 1T-TaS$_2$ and 2H-TaS$_2$ are 3.36\AA~and 3.316\AA ~respectively\cite{Mattheiss_PRB1973, Clerc_2004JoPCM}, which results in a moir\'e pattern with a periodicity of $L \sim 75$ unit cells.

Motivated by these recent developments, we develop a model to study the magnetic order in synthetic Kondo moir\'e superlattices. We consider honeycomb lattices for both layers and an extended Kondo interaction since the local stacking order changes within the moir\'e unit cell. In the limit of small twist angles, only a few sites are aligned perfectly on top of each other, corresponding to AA and AB stacking regions. For small twist angles, we calculate the Ruderman-Kittel-Kasuya-Yosida (RKKY) interaction\cite{Ruderman_PR1954, Yosida_PR1957, Kasuya_PTP1956} via exact diagonalization up to third nearest neighbor and perform Monte Carlo simulations on the effective spin model to obtain the ground state spin texture. Our main results are as follows: (i) we show that the sign and the magnitude of the RKKY interaction varies depending on the local stacking order. (ii) Our Monte Carlo simulations indicate that the AA stacking regions order antiferromagnetically (AFM), whereas the AB regions order ferromagnetically (FM). The remaining parts of the moir\'e unit cell form ferromagnetic chains that are coupled antiferromagnetically (FMC). (iii) The extent of these regions varies with the inverse decay length of the Kondo interaction, $\alpha$. For small $\alpha$, FMC dominates the majority of the moir\'e unit cell, whereas for large $\alpha$, FM, AFM, and FMC domains coexist.

We start by introducing the extended Kondo lattice Hamiltonian, which describes the interactions between localized magnetic moments with the conduction electrons in the moir\'e unit cell.
\begin{equation}
    H=-t\sum_{<ij>\sigma}(c_{ai\sigma}^{\dag}c_{bj\sigma} + {\rm H.c.})-\mu\sum_{is\sigma}c^\dagger_{is\sigma}c_{is\sigma}+\sum_{ij;\alpha \beta} J_K(r_{ij})
    c_{i\alpha}^\dagger \boldsymbol\sigma^{\alpha \beta} c_{i\beta} \cdot {\bf S}_j
\end{equation}
where $c^\dagger_{si\sigma}$ is the creation operator for conduction electrons in $s=\{a, b\}$ sublattice and ${\bf S}$ is the spin operator of the local moments. $r_{ij}$ represents the distance between two sites, $r_{ij} = |{\bf r}_i - {\bf r}_j|$. We consider the an extended Kondo coupling given by, 
\begin{eqnarray}
J_K(r_{ij}) &=& J_0 e^{-\alpha r_{ij}/a_0}~~{\rm for}~{r_{ij}\le \sqrt{3}a_0} \nonumber \\
&=&0~~{\rm otherwise.}
\label{eq:Hk}
\end{eqnarray}
where $a_0$ is the bond length on the honeycomb lattice and $\alpha$ is the inverse decay length. The cut-off in the exchange coupling for $r_{ij}>\sqrt{3}a_0$ is imposed for practical computational purposes\cite{Akram_NanoLett2024}. For the rest of the manuscript, we consider $\mu=0$, a semimetallic filling for the conduction electrons and use $J_0/t = 10^{-2}$ to stay in the perturbative regime. Extended Kondo interaction is essential for the topological phases in heavy fermions\cite{Dzero_ARCMP2016, Alexandrov_PRL2015, Ghazaryan_NJP2021, Lai_PNAF2018}. It has also been investigated for nematic\cite{Vijayvargia_arXiv2024} and higher angular momentum Kondo liquids\cite{Ghaemi_PRB2007, Ghaemi_PRB2008}.

RKKY interaction arises due to magnetic impurities interacting with conduction electrons which create net spin polarization. While the polarization propagates, it decays and undergoes Friedel oscillations. In Fig.~\ref{Fig1}(c), we present the spin polarization when the magnetic impurity is perfectly aligned on top of a conduction site (AA stacking). The polarization is antiferromagnetic to the magnetic impurity due to the nature of the Kondo exchange coupling. The coupling flips sign at its nearest neighbors (NN) due to the large Dirac momentum (short wavelength) Friedel oscillations, similar to the effect of magnetic impurities in graphene\cite{Black2010}. In Fig.~\ref{Fig1}(d), the magnetic impurity is placed at the center of the honeycomb (AB stacking). Note that the polarization is much smaller in this case since the local moment can only interact with conduction electrons at a distance $r = a_0$ and $J_K(r=a_0)\ll J_K(r=0)$. The interaction of a second impurity with the spin-polarization induced by the first impurity leads to an effective interaction between the two local moments, which is known as RKKY interaction. In the absence of spin-orbit coupling, the RKKY interaction has SU(2) symmetry and can be expressed as a Heisenberg interaction, $J_{ij}{\bf S}_{i} \cdot {\bf S}_{j}$. We estimate the RKKY coupling constant via exact diagonalization following the procedure described in Ref.~\citenum{Black2010}. We place two local moments at sites $i$ and $j$ and calculate the energy, including the interactions with the conduction electrons by aligning local moments parallel ($E_{FM}$) or antiparallel ($E_{AFM}$). The RKKY coupling is given by the energy difference, $J_{ij}=(E_{FM}-E_{AFM})/2S^2$. In the following, we consider unit spins, $S=1$. 

\begin{figure}[t]
\centering
   \includegraphics[width=0.9\textwidth]{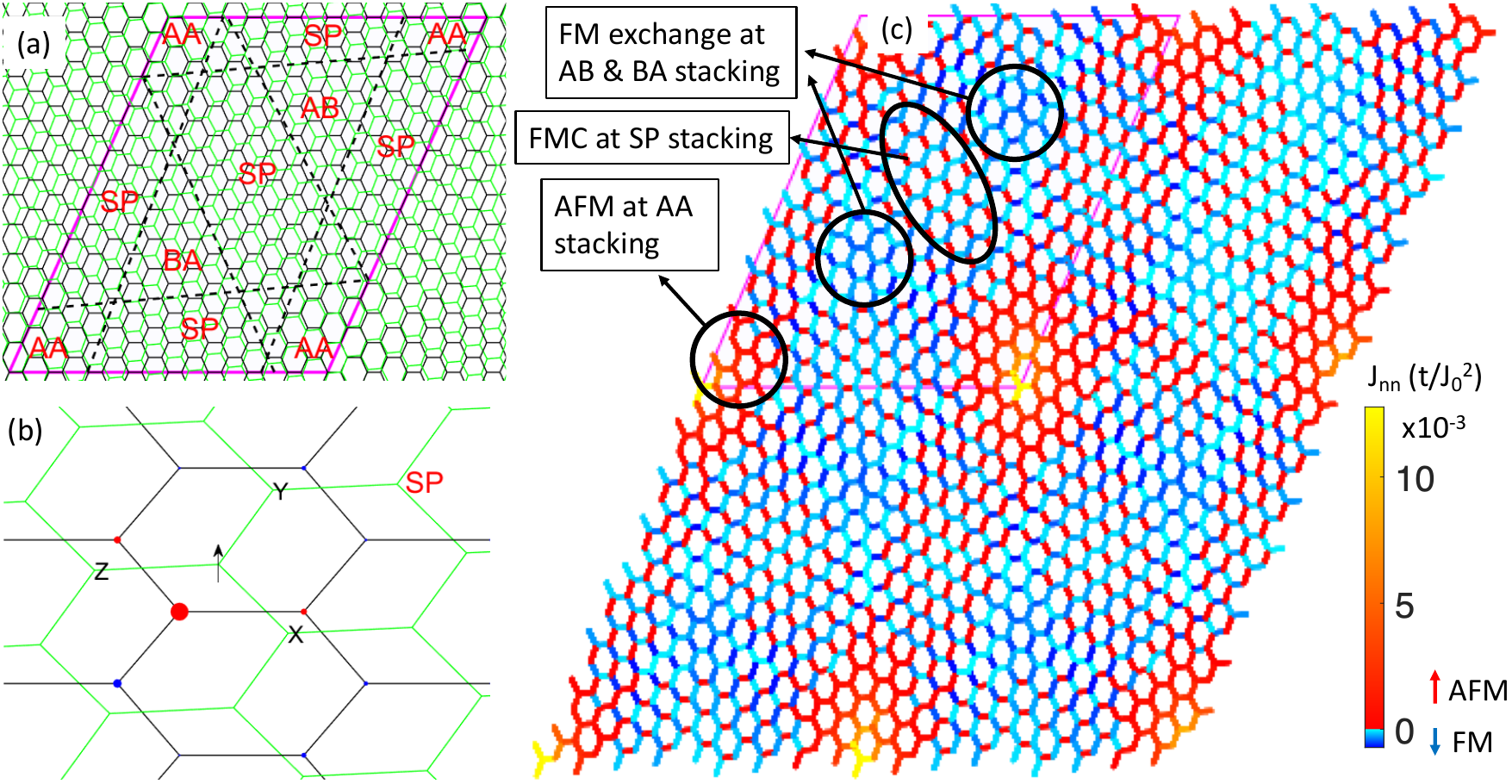}
    \caption{(a) Different stacking regions within a moir\'e unit cell: AA, AB, BA and saddle point (SP). (b) Conduction electron polarization induced by local moment for SP region for $\alpha=4.6$. (c) NN RKKY interaction within a 2x2 moir\'e unit cell. Note that the dominant interaction is FM (AFM) for AB and BA (AA) regions. The SP regions 2 of the 3 bonds are FM and the remaining bond is AFM which results in FM chains that are coupled antiferromagnetically.}
    \label{Fig2}
\end{figure}

There are two key factors in determining the sign and the magnitude of the RKKY interaction, and both depend on the local stacking pattern. The magnitude of the RKKY interaction is primarily determined by the magnitude of $r_{ij}$. Since $J_K(r_{ij})$ decays exponentially with $r_{ij}$ (eq.~\ref{eq:Hk}), the RKKY interaction which is proportional to $[J_K(r_{ij})]^2$ is overall larger in AA stacking regions where $r_{ij}/a_0\ll1$ compared to AB stacking where $r_{ij}/a_0\sim1$ for one of the magnetic impurity. The sign of the interaction depends on whether the local moments primarily couple to the same or different conduction sites. Coupling to the same conduction site results in an FM RKKY interaction. Yet, if they couple to NN sites, the sign of the conduction electron polarization flips between the NN sites due to Friedel oscillations as discussed above and the RKKY interaction is AFM in this case. To illustrate these effects, we present the RKKY interaction as a function of $\alpha$ for AA stacking and AB stacking for the `untwisted' system
in Fig.~\ref{Fig1}(e) and (f). For $\alpha \gg 1$, we recover the limit of local Kondo interaction where the exchange is only significant for $r_{ij}/a_0\ll 1$. In particular, the AA stacking shows AFM NN and FM next nearest neighbor (NNN) interactions. These results are in agreement with previous estimations for RKKY interaction in graphene\cite{Black2010}. On the contrary, the exchange constants are significantly smaller for the AB stacking region since $J_{K}(r=a_0)$ diminishes for $\alpha \gg 1$. However, unlike AA stacking, the NN exchange is FM since the local moments couple to the same conduction site. 

For small $\alpha$, the behavior of the exchange couplings is more complex due to multiple channels of Kondo interaction interfering with each other\cite{Ahamed_PRB2018}. In particular, we find that the sign of the NN interaction changes to FM for AA stacking for $\alpha < 2$ and several other exchange couplings seem to diverge for $\alpha <1$. In order to keep $J_K(r= \sqrt{3}a_0)/J_K(r=0)$ small and justify the cut-off at $r > \sqrt{3}a_0$, we  only consider $\alpha \ge 3$ for the reminder of the paper. To the best of our knowledge, the decay length dependence of the interlayer Kondo interaction in vdW materials has not been studied extensively. Still, we anticipate that $3 \le \alpha \le 10$ provides a regime that covers a large parameter space.

\begin{figure}
\centering
   \includegraphics[width=1\textwidth]{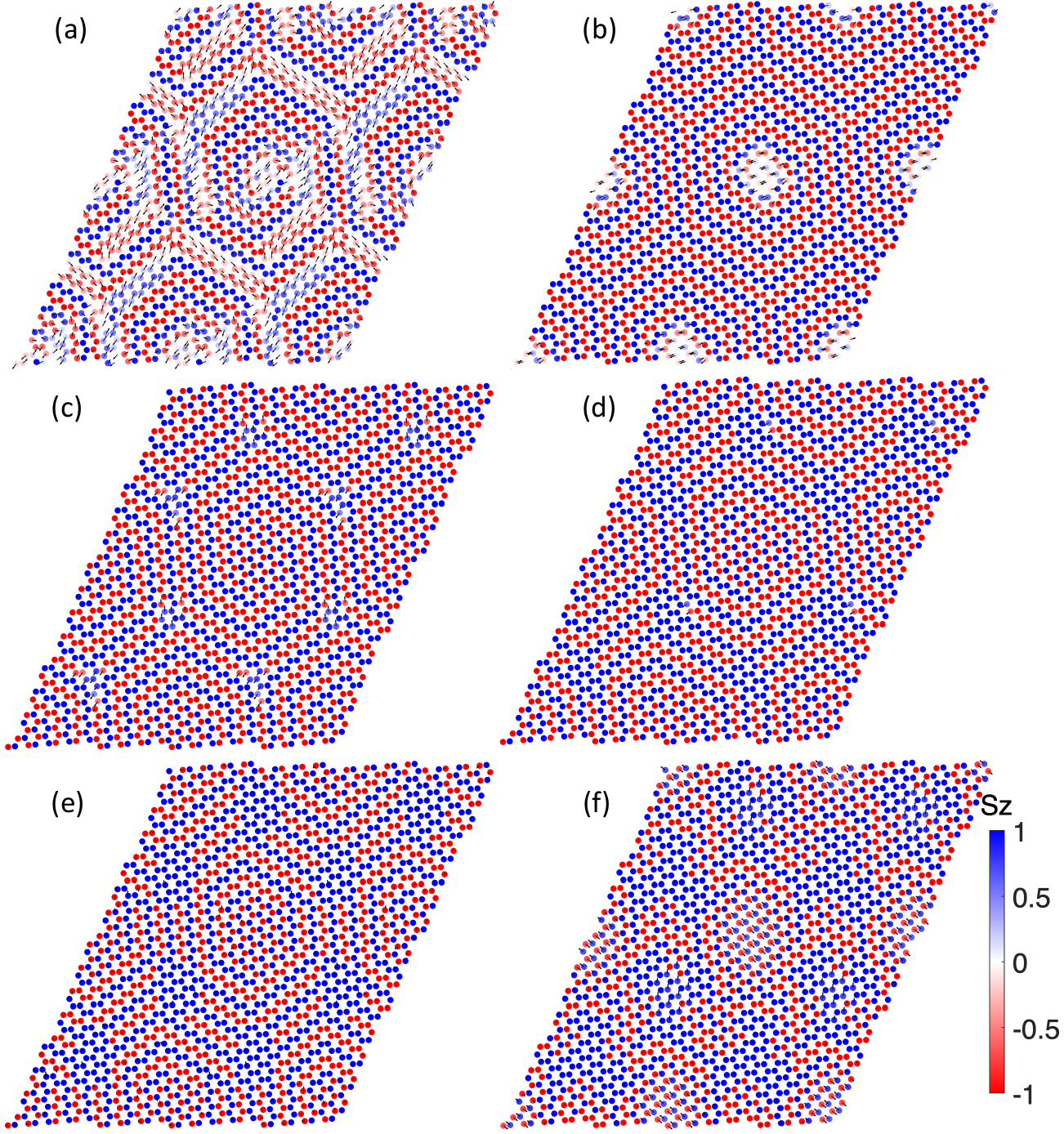}
    \caption{Monte Carlo snapshots for the ground state spin textures for $\alpha=3,3.2,4.3,4.6,5,10$ for (a)-(f) respectively. For $\alpha <4$, the ground state primarily consists of ferromagnetic chains that are coupled antiferromagnetically with a small AFM region at the AA stacking. For $\alpha>4$, FM (AFM) regions grow at AB and BA (AA) stacking regions within the moir\'e unit cell. The color plot shows the z-component of the magnetization, whereas the arrows show the in-plane component.}
    \label{Fig3}
\end{figure}


For small twist angles, the local stacking within a moir\'e unit cell can be described by a translational shift, ${\bf r}(\bf{R}) \simeq  \theta \hat{z} \times {\bf R}$ where ${\bf R}$ and ${\bf r}$ are the position within the moir\'e unit cell and translational shift respectively \cite{Akram_NanoLett2021, Akram_NanoLett2024}. Therefore, a moir\'e unit cell with a large periodicity can be considered as a superposition of all possible stacking orders\cite{Akram_PRB2021}. We label the different regions in the moir\'e unit cell by the primary stacking orders\cite{McGilly_NatNano2020}, AA, AB, BA, and saddle point stacking (SP) in Fig.~\ref{Fig2}(a). These regions exhibit different types of RKKY interactions. A local moment placed in AA or AB stacking regions polarizes the conduction electrons similar to the untwisted case as depicted in Fig.~\ref{Fig1}(c) and (d). As a result, the effective magnetic interactions in these regions are similar to the untwisted case but superimposed in real space. The polarization induced by a local moment in the saddle point (SP) region is shown in Fig.~\ref{Fig2}(b). It is noteworthy to emphasize that among the three NN sites, two of them couple primarily to the same conduction site and therefore exhibit FM exchange. However, for the other bonds, the local moments couple to the different conduction sites, which results in AFM NN coupling. This leads to FM chains (FMC) in the SP region, as discussed below. We summarize these results in Fig.~\ref{Fig2}(c), where we present the NN exchanges for all the bonds within a $2\times2$ moir\'e unit cell for $\alpha=4.6$. We consider a twist angle $\theta = 4^\circ$ with $N=394$ unit cells for the remainder of the article, but our results are independent of the twist angle for small $\theta$.

In order to determine the magnetic ground state of the moir\'e Kondo lattice, we derive the RKKY exchange couplings up to the third NN and perform classical Monte Carlo studies via the Metropolis algorithm. We include a small single site anisotropy term, $-J_A (S_i^z)^2$, which can naturally arise in real materials due to spin-orbit coupling\cite{Banerjee_PRX2014}, to break the SU(2) symmetry of the model. We pick $J_A$ much less than the smallest exchange coupling.  Starting with random initial conditions, we perform 120,000 Monte Carlo updates at each temperature. We determine the ground state at $T/J_{\rm max} = 10^{-4}$ where $J_{\rm max}$ is the maximum exchange coupling. Fig.~\ref{Fig3} depicts the Monte Carlo snapshots of the ground state spin configurations for different values of $\alpha$. We demonstrate that for $\alpha <4$, the ground state primarily consists of FMC throughout the moir\'e unit cell and a small fraction of AFM around the AA stacking regions. For $\alpha >4$, the FMC starts to shrink and FM (AFM) regions grow around AB and BA (AA) regions. FMC regions almost fully disappear as depicted in Fig.~\ref{Fig3}(f) for $\alpha=10$. Note that for such values of $\alpha$, RKKY coupling essentially vanishes except for the AA stacking regions.

\begin{figure}
\centering
   \includegraphics[width=1\textwidth]{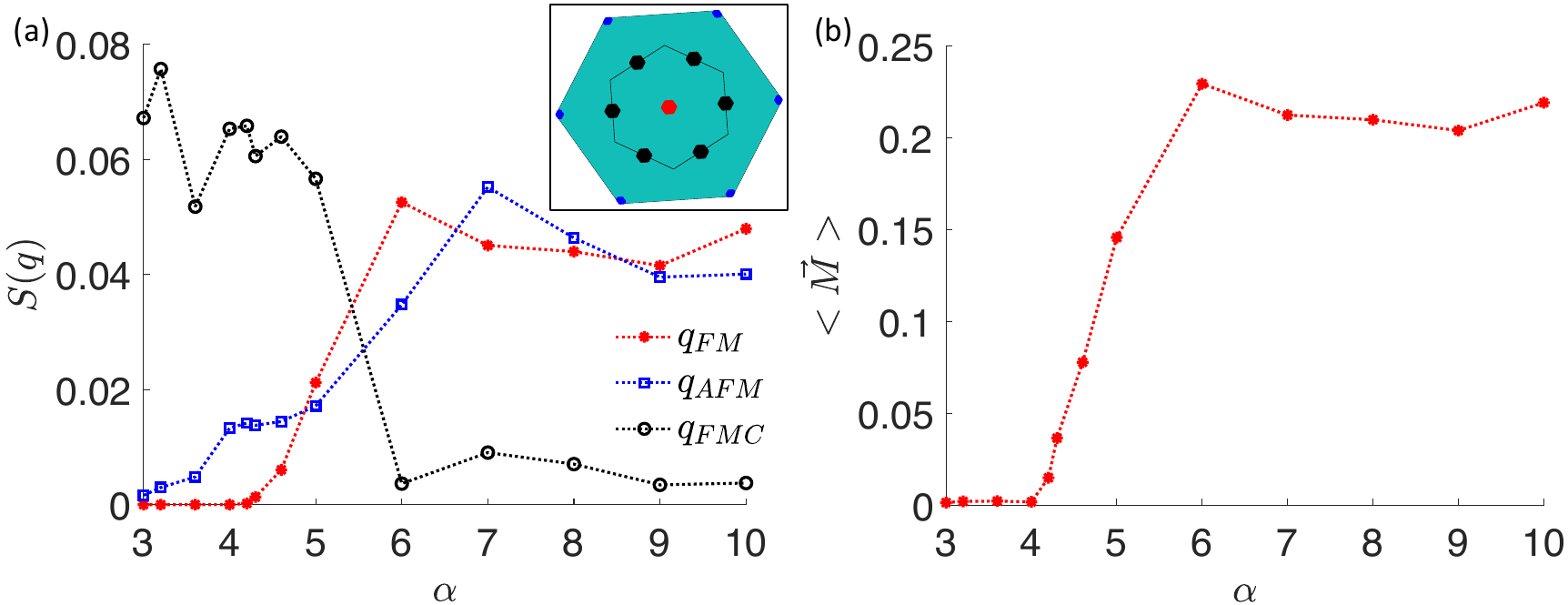}
    \caption{(a) Static spin structure factor S({\bf q}) as a function of $\alpha$, for $ {\bf q}=(0,0)$ (red), $ {\bf {q}}=\pm{\bf{b}}_1, \pm{\bf{b}}_2, \pm({\bf{b}}_1+{\bf{b}}_2)$ (blue) and $ {\bf{q} }=\pm{\bf{b}}_1/2, \pm{\bf{b}}_2/2, \pm({\bf{b}}_1+{\bf{b}}_2)/2$ (black) corresponding to FM, AFM and FMC order respectively. The inset shows the position of these wave vectors in the extended Brillouin zone. (b) Net magnetization as a function of $\alpha$. Note that the ${\bf{q}}=(0,0)$ component of the spin structure factor is the square of magnetization.}
    \label{Fig4}
\end{figure}
In order to quantify these trends, we calculate the magnetization, $\langle M^\alpha \rangle =\langle\frac{1}{N}\sum_i S^\alpha_i \rangle$,
and the static spin structure factor    $S({\textbf{q}})=(1/N^2)\langle\sum_{\beta=x,y,x} |\sum_iS_i^{\beta}e^{-i\textbf{q}.\textbf{r}_i}|^2 \rangle$ of the ground state. $S({\textbf{q}})$ exhibits peaks at ${\bf q} = (0,0),~\{\pm {\bf b}_{1(2)}/2, \pm({\bf b}_1+{\bf b}_2)/2\},~\{\pm{\bf b}_{1(2)}, \pm({\bf b}_1+{\bf b}_2) \}$, where ${\bf b}_{1/2}$ are the reciprocal lattice vectors, corresponding to FM, FMC and AFM orders respectively (see Supporting Information for details). We present the amplitude of these peaks in Fig.~\ref{Fig4}(a). The  ${\bf q} = (0,0)$ component of $S({\textbf{q}})$ is the square of the magnetization as shown in Fig.~\ref{Fig4}(b). The spin structure factor also shows that the FMC is the primary order for $\alpha<6$. For $\alpha > 6$, FMC regions shrink and AFM, FM regions grow rapidly. As shown in Fig.~\ref{Fig4}(b), the magnetization grows for $\alpha>4$ and saturates at about $M/S= 0.2-0.25$.

Experimental detection of complex magnetic textures in the 2D limit is challenging. Although a variety of non-coplanar magnetic structures have been predicted in moir\'e superlattices of vdW magnets\cite{Tong_ACS2018, Hejazi_PNAS2020, Hejazi_PRB2021, Akram_PRB2021, Akram_NanoLett2021, Tong_PRR2021, Xiao_PRR2021, Ghader_CommPhys2022, Zheng_AFM2023, Kim_NanoLett2023, Fumega_2DMAtt2023, Nica_npjQM2023, Vijayvargia_PRR2023}, only few of these phases have been observed experimentally by magnetic circular dichroism\cite{Xu_NatNano2021}, magneto-optical Kerr effect\cite{Xie_NatPhys2022} and diamond NV magnetometry\cite{Song_Science2021}. In particular, we expect that these techniques should be able to detect the FM domains.

In conclusion, we presented a theoretical framework for magnetic order in synthetic moir\'e Kondo lattices consisting of Mott insulator and metal vdW layers. We showed that the stacking-dependent Kondo interaction can give rise to various types of magnetic order. In particular, we demonstrated that the decay length of the Kondo interaction can control the magnetic textures. Interesting future directions include realistic simulations of material platforms, investigating the effects of pressure and gating, and the competition between heavy fermion formation and magnetism in these systems.

\section{Associated Content}

\noindent {\bf Supporting Information}

\noindent Details of the static spin structure factor calculation (pdf).

\section{Author Information}
\noindent {\bf Corresponding Author:}
Onur Erten -- Email: onur.erten@asu.edu

\noindent {\bf Author Contributions:}
M. A. Keskiner have performed the exact diagonalization and Monte Carlo simulations. M. \"O. Oktel, Pouyan Ghaemi and Onur Erten have conceptualized the research. All authors have analyzed the results and contributed to writing the manuscript.

\noindent {\bf Notes:} The authors declare no competing financial interest.

\section{Acknowledgements}
We thank Cemal Yalabik and Erik Henriksen for fruitful discussions. OE acknowledge support from National Science Foundation Award No. DMR 2234352. MAK and MOO are partially supported by TUBITAK 1001 grant No 122F346. PG acknowledges support from NSF DMR-2130544 and infrastructural support from NSF HRD-2112550 (NSF CREST Center IDEALS).

\providecommand{\latin}[1]{#1}
\makeatletter
\providecommand{\doi}
  {\begingroup\let\do\@makeother\dospecials
  \catcode`\{=1 \catcode`\}=2 \doi@aux}
\providecommand{\doi@aux}[1]{\endgroup\texttt{#1}}
\makeatother
\providecommand*\mcitethebibliography{\thebibliography}
\csname @ifundefined\endcsname{endmcitethebibliography}  {\let\endmcitethebibliography\endthebibliography}{}


\begin{mcitethebibliography}{49}
\providecommand*\natexlab[1]{#1}
\providecommand*\mciteSetBstSublistMode[1]{}
\providecommand*\mciteSetBstMaxWidthForm[2]{}
\providecommand*\mciteBstWouldAddEndPuncttrue
  {\def\EndOfBibitem{\unskip.}}
\providecommand*\mciteBstWouldAddEndPunctfalse
  {\let\EndOfBibitem\relax}
\providecommand*\mciteSetBstMidEndSepPunct[3]{}
\providecommand*\mciteSetBstSublistLabelBeginEnd[3]{}
\providecommand*\EndOfBibitem{}
\mciteSetBstSublistMode{f}
\mciteSetBstMaxWidthForm{subitem}{(\alph{mcitesubitemcount})}
\mciteSetBstSublistLabelBeginEnd
  {\mcitemaxwidthsubitemform\space}
  {\relax}
  {\relax}

\bibitem[Stewart(1984)]{Stewart_RMP1984}
Stewart,~G.~R. Heavy-fermion systems. \emph{Rev. Mod. Phys.} \textbf{1984}, \emph{56}, 755--787\relax
\mciteBstWouldAddEndPuncttrue
\mciteSetBstMidEndSepPunct{\mcitedefaultmidpunct}
{\mcitedefaultendpunct}{\mcitedefaultseppunct}\relax
\EndOfBibitem
\bibitem[Coleman(2015)]{Coleman_book}
Coleman,~P. \emph{Introduction to Many-Body Physics}; Cambridge University Press, 2015\relax
\mciteBstWouldAddEndPuncttrue
\mciteSetBstMidEndSepPunct{\mcitedefaultmidpunct}
{\mcitedefaultendpunct}{\mcitedefaultseppunct}\relax
\EndOfBibitem
\bibitem[Wirth and Steglich(2016)Wirth, and Steglich]{Wirth_NatRevMat2016}
Wirth,~S.; Steglich,~F. Exploring heavy fermions from macroscopic to microscopic length scales. \emph{Nature Reviews Materials} \textbf{2016}, \emph{1}, 16051\relax
\mciteBstWouldAddEndPuncttrue
\mciteSetBstMidEndSepPunct{\mcitedefaultmidpunct}
{\mcitedefaultendpunct}{\mcitedefaultseppunct}\relax
\EndOfBibitem
\bibitem[Ayani \latin{et~al.}(2023)Ayani, Pisarra, Ibarburu, Garnica, Miranda, Calleja, Martin, and Vazquez~de Parga]{Ayani_Small2023}
Ayani,~C.~G.; Pisarra,~M.; Ibarburu,~I.~M.; Garnica,~M.; Miranda,~R.; Calleja,~F.; Martin,~F.; Vazquez~de Parga,~A.~L. Probing the Phase Transition to a Coherent 2D {K}ondo Lattice. \emph{Small} \textbf{2023}, 2303275\relax
\mciteBstWouldAddEndPuncttrue
\mciteSetBstMidEndSepPunct{\mcitedefaultmidpunct}
{\mcitedefaultendpunct}{\mcitedefaultseppunct}\relax
\EndOfBibitem
\bibitem[Jin and Knolle(2021)Jin, and Knolle]{Jin_PRB2021}
Jin,~H.-K.; Knolle,~J. Flat and correlated plasmon bands in $\text{graphene}/\alpha$-$\mathrm{RuCl}_{3}$ heterostructures. \emph{Phys. Rev. B} \textbf{2021}, \emph{104}, 045140\relax
\mciteBstWouldAddEndPuncttrue
\mciteSetBstMidEndSepPunct{\mcitedefaultmidpunct}
{\mcitedefaultendpunct}{\mcitedefaultseppunct}\relax
\EndOfBibitem
\bibitem[Mashhadi \latin{et~al.}(2019)Mashhadi, Kim, Kim, Weber, Taniguchi, Watanabe, Park, Lotsch, Smet, Burghard, \latin{et~al.} others]{mashhadi2019}
Mashhadi,~S.; Kim,~Y.; Kim,~J.; Weber,~D.; Taniguchi,~T.; Watanabe,~K.; Park,~N.; Lotsch,~B.; Smet,~J.~H.; Burghard,~M.; others Spin-split band hybridization in graphene proximitized with $\alpha$-RuCl3 nanosheets. \emph{Nano letters} \textbf{2019}, \emph{19}, 4659--4665\relax
\mciteBstWouldAddEndPuncttrue
\mciteSetBstMidEndSepPunct{\mcitedefaultmidpunct}
{\mcitedefaultendpunct}{\mcitedefaultseppunct}\relax
\EndOfBibitem
\bibitem[Zhou \latin{et~al.}(2019)Zhou, Balgley, Lampen-Kelley, Yan, Mandrus, and Henriksen]{zhou2019}
Zhou,~B.; Balgley,~J.; Lampen-Kelley,~P.; Yan,~J.-Q.; Mandrus,~D.~G.; Henriksen,~E.~A. Evidence for charge transfer and proximate magnetism in graphene--$\alpha$- RuCl 3 heterostructures. \emph{Physical Review B} \textbf{2019}, \emph{100}, 165426\relax
\mciteBstWouldAddEndPuncttrue
\mciteSetBstMidEndSepPunct{\mcitedefaultmidpunct}
{\mcitedefaultendpunct}{\mcitedefaultseppunct}\relax
\EndOfBibitem
\bibitem[Biswas \latin{et~al.}(2019)Biswas, Li, Winter, Knolle, and Valent{\'\i}]{biswas2019}
Biswas,~S.; Li,~Y.; Winter,~S.~M.; Knolle,~J.; Valent{\'\i},~R. Electronic Properties of $\alpha$- RuCl 3 in proximity to graphene. \emph{Physical Review Letters} \textbf{2019}, \emph{123}, 237201\relax
\mciteBstWouldAddEndPuncttrue
\mciteSetBstMidEndSepPunct{\mcitedefaultmidpunct}
{\mcitedefaultendpunct}{\mcitedefaultseppunct}\relax
\EndOfBibitem
\bibitem[Rizzo \latin{et~al.}(2020)Rizzo, Jessen, Sun, Ruta, Zhang, Yan, Xian, McLeod, Berkowitz, Watanabe, Taniguchi, Nagler, Mandrus, Rubio, Fogler, Millis, Hone, Dean, and Basov]{Rizzo_NanoLett2020}
Rizzo,~D.~J. \latin{et~al.}  Charge-Transfer Plasmon Polaritons at Graphene/$\alpha$-RuCl3 Interfaces. \emph{Nano Letters} \textbf{2020}, \emph{20}, 8438--8445\relax
\mciteBstWouldAddEndPuncttrue
\mciteSetBstMidEndSepPunct{\mcitedefaultmidpunct}
{\mcitedefaultendpunct}{\mcitedefaultseppunct}\relax
\EndOfBibitem
\bibitem[Gerber \latin{et~al.}(2020)Gerber, Yao, Arias, and Kim]{gerber2020}
Gerber,~E.; Yao,~Y.; Arias,~T.~A.; Kim,~E.-A. Ab Initio Mismatched Interface Theory of Graphene on $\alpha$- RuCl 3: Doping and Magnetism. \emph{Physical Review Letters} \textbf{2020}, \emph{124}, 106804\relax
\mciteBstWouldAddEndPuncttrue
\mciteSetBstMidEndSepPunct{\mcitedefaultmidpunct}
{\mcitedefaultendpunct}{\mcitedefaultseppunct}\relax
\EndOfBibitem
\bibitem[Leeb \latin{et~al.}(2021)Leeb, Polyudov, Mashhadi, Biswas, Valent{\'\i}, Burghard, and Knolle]{leeb_PRL2021}
Leeb,~V.; Polyudov,~K.; Mashhadi,~S.; Biswas,~S.; Valent{\'\i},~R.; Burghard,~M.; Knolle,~J. Anomalous quantum oscillations in a heterostructure of graphene on a proximate quantum spin liquid. \emph{Physical Review Letters} \textbf{2021}, \emph{126}, 097201\relax
\mciteBstWouldAddEndPuncttrue
\mciteSetBstMidEndSepPunct{\mcitedefaultmidpunct}
{\mcitedefaultendpunct}{\mcitedefaultseppunct}\relax
\EndOfBibitem
\bibitem[Balgley \latin{et~al.}(2022)Balgley, Butler, Biswas, Ge, Lagasse, Taniguchi, Watanabe, Cothrine, Mandrus, Velasco~Jr, \latin{et~al.} others]{balgley2022}
Balgley,~J.; Butler,~J.; Biswas,~S.; Ge,~Z.; Lagasse,~S.; Taniguchi,~T.; Watanabe,~K.; Cothrine,~M.; Mandrus,~D.~G.; Velasco~Jr,~J.; others Ultrasharp Lateral p--n Junctions in Modulation-Doped Graphene. \emph{Nano Letters} \textbf{2022}, \emph{22}, 4124--4130\relax
\mciteBstWouldAddEndPuncttrue
\mciteSetBstMidEndSepPunct{\mcitedefaultmidpunct}
{\mcitedefaultendpunct}{\mcitedefaultseppunct}\relax
\EndOfBibitem
\bibitem[Shi and MacDonald(2023)Shi, and MacDonald]{shi2023magnetic}
Shi,~J.; MacDonald,~A. Magnetic states of graphene proximitized Kitaev materials. \emph{Physical Review B} \textbf{2023}, \emph{108}, 064401\relax
\mciteBstWouldAddEndPuncttrue
\mciteSetBstMidEndSepPunct{\mcitedefaultmidpunct}
{\mcitedefaultendpunct}{\mcitedefaultseppunct}\relax
\EndOfBibitem
\bibitem[Liu \latin{et~al.}(2016)Liu, Weiss, Duan, Cheng, Huang, and Duan]{Liu_NatRevMat2016}
Liu,~Y.; Weiss,~N.~O.; Duan,~X.; Cheng,~H.-C.; Huang,~Y.; Duan,~X. Van der Waals heterostructures and devices. \emph{Nature Reviews Materials} \textbf{2016}, \emph{1}, 16042\relax
\mciteBstWouldAddEndPuncttrue
\mciteSetBstMidEndSepPunct{\mcitedefaultmidpunct}
{\mcitedefaultendpunct}{\mcitedefaultseppunct}\relax
\EndOfBibitem
\bibitem[He \latin{et~al.}(2021)He, Zhou, Ye, Cho, Jeong, Meng, and Wang]{He_ACSNano2021}
He,~F.; Zhou,~Y.; Ye,~Z.; Cho,~S.-H.; Jeong,~J.; Meng,~X.; Wang,~Y. Moir{\'e} Patterns in 2D Materials: A Review. \emph{ACS Nano} \textbf{2021}, \emph{15}, 5944--5958\relax
\mciteBstWouldAddEndPuncttrue
\mciteSetBstMidEndSepPunct{\mcitedefaultmidpunct}
{\mcitedefaultendpunct}{\mcitedefaultseppunct}\relax
\EndOfBibitem
\bibitem[Mattheiss(1973)]{Mattheiss_PRB1973}
Mattheiss,~L.~F. Band Structures of Transition-Metal-Dichalcogenide Layer Compounds. \emph{Phys. Rev. B} \textbf{1973}, \emph{8}, 3719--3740\relax
\mciteBstWouldAddEndPuncttrue
\mciteSetBstMidEndSepPunct{\mcitedefaultmidpunct}
{\mcitedefaultendpunct}{\mcitedefaultseppunct}\relax
\EndOfBibitem
\bibitem[Clerc \latin{et~al.}(2004)Clerc, Bovet, Berger, Despont, Koitzsch, Gallus, Patthey, Shi, Krempasky, Garnier, and Aebi]{Clerc_2004JoPCM}
Clerc,~F.; Bovet,~M.; Berger,~H.; Despont,~L.; Koitzsch,~C.; Gallus,~O.; Patthey,~L.; Shi,~M.; Krempasky,~J.; Garnier,~M.~G.; Aebi,~P. Spin–orbit splitting in the valence bands of 1T-TaS2 and 1T-TaSe2. \emph{Journal of Physics: Condensed Matter} \textbf{2004}, \emph{16}, 3271\relax
\mciteBstWouldAddEndPuncttrue
\mciteSetBstMidEndSepPunct{\mcitedefaultmidpunct}
{\mcitedefaultendpunct}{\mcitedefaultseppunct}\relax
\EndOfBibitem
\bibitem[Ruderman and Kittel(1954)Ruderman, and Kittel]{Ruderman_PR1954}
Ruderman,~M.~A.; Kittel,~C. Indirect Exchange Coupling of Nuclear Magnetic Moments by Conduction Electrons. \emph{Phys. Rev.} \textbf{1954}, \emph{96}, 99--102\relax
\mciteBstWouldAddEndPuncttrue
\mciteSetBstMidEndSepPunct{\mcitedefaultmidpunct}
{\mcitedefaultendpunct}{\mcitedefaultseppunct}\relax
\EndOfBibitem
\bibitem[Yosida(1957)]{Yosida_PR1957}
Yosida,~K. Magnetic Properties of {Cu-Mn} Alloys. \emph{Phys. Rev.} \textbf{1957}, \emph{106}, 893--898\relax
\mciteBstWouldAddEndPuncttrue
\mciteSetBstMidEndSepPunct{\mcitedefaultmidpunct}
{\mcitedefaultendpunct}{\mcitedefaultseppunct}\relax
\EndOfBibitem
\bibitem[Kasuya(1956)]{Kasuya_PTP1956}
Kasuya,~T. {A Theory of Metallic Ferro- and Antiferromagnetism on Zener's Model}. \emph{Progress of Theoretical Physics} \textbf{1956}, \emph{16}, 45--57\relax
\mciteBstWouldAddEndPuncttrue
\mciteSetBstMidEndSepPunct{\mcitedefaultmidpunct}
{\mcitedefaultendpunct}{\mcitedefaultseppunct}\relax
\EndOfBibitem
\bibitem[Akram \latin{et~al.}(2024)Akram, Kapeghian, Das, Valent{\'i}, Botana, and Erten]{Akram_NanoLett2024}
Akram,~M.; Kapeghian,~J.; Das,~J.; Valent{\'i},~R.; Botana,~A.~S.; Erten,~O. Theory of Moir{\'e} Magnetism in Twisted Bilayer $\alpha$-RuCl3. \emph{Nano Letters} \textbf{2024}, \emph{24}, 890--896\relax
\mciteBstWouldAddEndPuncttrue
\mciteSetBstMidEndSepPunct{\mcitedefaultmidpunct}
{\mcitedefaultendpunct}{\mcitedefaultseppunct}\relax
\EndOfBibitem
\bibitem[Dzero \latin{et~al.}(2016)Dzero, Xia, Galitski, and Coleman]{Dzero_ARCMP2016}
Dzero,~M.; Xia,~J.; Galitski,~V.; Coleman,~P. Topological Kondo Insulators. \emph{Annual Review of Condensed Matter Physics} \textbf{2016}, \emph{7}, 249--280\relax
\mciteBstWouldAddEndPuncttrue
\mciteSetBstMidEndSepPunct{\mcitedefaultmidpunct}
{\mcitedefaultendpunct}{\mcitedefaultseppunct}\relax
\EndOfBibitem
\bibitem[Alexandrov \latin{et~al.}(2015)Alexandrov, Coleman, and Erten]{Alexandrov_PRL2015}
Alexandrov,~V.; Coleman,~P.; Erten,~O. Kondo Breakdown in Topological Kondo Insulators. \emph{Phys. Rev. Lett.} \textbf{2015}, \emph{114}, 177202\relax
\mciteBstWouldAddEndPuncttrue
\mciteSetBstMidEndSepPunct{\mcitedefaultmidpunct}
{\mcitedefaultendpunct}{\mcitedefaultseppunct}\relax
\EndOfBibitem
\bibitem[Ghazaryan \latin{et~al.}(2021)Ghazaryan, Nica, Erten, and Ghaemi]{Ghazaryan_NJP2021}
Ghazaryan,~A.; Nica,~E.~M.; Erten,~O.; Ghaemi,~P. Shadow surface states in topological Kondo insulators. \emph{New Journal of Physics} \textbf{2021}, \emph{23}, 123042\relax
\mciteBstWouldAddEndPuncttrue
\mciteSetBstMidEndSepPunct{\mcitedefaultmidpunct}
{\mcitedefaultendpunct}{\mcitedefaultseppunct}\relax
\EndOfBibitem
\bibitem[Lai \latin{et~al.}(2018)Lai, Grefe, Paschen, and Si]{Lai_PNAF2018}
Lai,~H.-H.; Grefe,~S.~E.; Paschen,~S.; Si,~Q. Weyl–Kondo semimetal in heavy-fermion systems. \emph{Proceedings of the National Academy of Sciences} \textbf{2018}, \emph{115}, 93--97\relax
\mciteBstWouldAddEndPuncttrue
\mciteSetBstMidEndSepPunct{\mcitedefaultmidpunct}
{\mcitedefaultendpunct}{\mcitedefaultseppunct}\relax
\EndOfBibitem
\bibitem[Vijayvargia and Erten(2024)Vijayvargia, and Erten]{Vijayvargia_arXiv2024}
Vijayvargia,~A.; Erten,~O. Nematic heavy fermions and coexisting magnetic order in CeSiI. 2024\relax
\mciteBstWouldAddEndPuncttrue
\mciteSetBstMidEndSepPunct{\mcitedefaultmidpunct}
{\mcitedefaultendpunct}{\mcitedefaultseppunct}\relax
\EndOfBibitem
\bibitem[Ghaemi and Senthil(2007)Ghaemi, and Senthil]{Ghaemi_PRB2007}
Ghaemi,~P.; Senthil,~T. Higher angular momentum Kondo liquids. \emph{Phys. Rev. B} \textbf{2007}, \emph{75}, 144412\relax
\mciteBstWouldAddEndPuncttrue
\mciteSetBstMidEndSepPunct{\mcitedefaultmidpunct}
{\mcitedefaultendpunct}{\mcitedefaultseppunct}\relax
\EndOfBibitem
\bibitem[Ghaemi \latin{et~al.}(2008)Ghaemi, Senthil, and Coleman]{Ghaemi_PRB2008}
Ghaemi,~P.; Senthil,~T.; Coleman,~P. Angle-dependent quasiparticle weights in correlated metals. \emph{Phys. Rev. B} \textbf{2008}, \emph{77}, 245108\relax
\mciteBstWouldAddEndPuncttrue
\mciteSetBstMidEndSepPunct{\mcitedefaultmidpunct}
{\mcitedefaultendpunct}{\mcitedefaultseppunct}\relax
\EndOfBibitem
\bibitem[Black-Schaffer(2010)]{Black2010}
Black-Schaffer,~A.~M. RKKY coupling in graphene. \emph{Phys. Rev. B} \textbf{2010}, \emph{81}, 205416\relax
\mciteBstWouldAddEndPuncttrue
\mciteSetBstMidEndSepPunct{\mcitedefaultmidpunct}
{\mcitedefaultendpunct}{\mcitedefaultseppunct}\relax
\EndOfBibitem
\bibitem[Ahamed \latin{et~al.}(2018)Ahamed, Moessner, and Erten]{Ahamed_PRB2018}
Ahamed,~S.; Moessner,~R.; Erten,~O. Why rare-earth ferromagnets are so rare: Insights from the $p$-wave Kondo model. \emph{Phys. Rev. B} \textbf{2018}, \emph{98}, 054420\relax
\mciteBstWouldAddEndPuncttrue
\mciteSetBstMidEndSepPunct{\mcitedefaultmidpunct}
{\mcitedefaultendpunct}{\mcitedefaultseppunct}\relax
\EndOfBibitem
\bibitem[Akram \latin{et~al.}(2021)Akram, LaBollita, Dey, Kapeghian, Erten, and Botana]{Akram_NanoLett2021}
Akram,~M.; LaBollita,~H.; Dey,~D.; Kapeghian,~J.; Erten,~O.; Botana,~A.~S. Moir{\'e} Skyrmions and Chiral Magnetic Phases in Twisted CrX3 (X = I, Br, and Cl) Bilayers. \emph{Nano Letters} \textbf{2021}, \emph{21}, 6633--6639\relax
\mciteBstWouldAddEndPuncttrue
\mciteSetBstMidEndSepPunct{\mcitedefaultmidpunct}
{\mcitedefaultendpunct}{\mcitedefaultseppunct}\relax
\EndOfBibitem
\bibitem[Akram and Erten(2021)Akram, and Erten]{Akram_PRB2021}
Akram,~M.; Erten,~O. Skyrmions in twisted van der Waals magnets. \emph{Phys. Rev. B} \textbf{2021}, \emph{103}, L140406\relax
\mciteBstWouldAddEndPuncttrue
\mciteSetBstMidEndSepPunct{\mcitedefaultmidpunct}
{\mcitedefaultendpunct}{\mcitedefaultseppunct}\relax
\EndOfBibitem
\bibitem[McGilly \latin{et~al.}(2020)McGilly, Kerelsky, Finney, Shapovalov, Shih, Ghiotto, Zeng, Moore, Wu, Bai, Watanabe, Taniguchi, Stengel, Zhou, Hone, Zhu, Basov, Dean, Dreyer, and Pasupathy]{McGilly_NatNano2020}
McGilly,~L.~J. \latin{et~al.}  Visualization of moir{\'e} superlattices. \emph{Nature Nanotechnology} \textbf{2020}, \emph{15}, 580--584\relax
\mciteBstWouldAddEndPuncttrue
\mciteSetBstMidEndSepPunct{\mcitedefaultmidpunct}
{\mcitedefaultendpunct}{\mcitedefaultseppunct}\relax
\EndOfBibitem
\bibitem[Banerjee \latin{et~al.}(2014)Banerjee, Rowland, Erten, and Randeria]{Banerjee_PRX2014}
Banerjee,~S.; Rowland,~J.; Erten,~O.; Randeria,~M. Enhanced Stability of Skyrmions in Two-Dimensional Chiral Magnets with Rashba Spin-Orbit Coupling. \emph{Phys. Rev. X} \textbf{2014}, \emph{4}, 031045\relax
\mciteBstWouldAddEndPuncttrue
\mciteSetBstMidEndSepPunct{\mcitedefaultmidpunct}
{\mcitedefaultendpunct}{\mcitedefaultseppunct}\relax
\EndOfBibitem
\bibitem[Tong \latin{et~al.}(2018)Tong, Liu, Xiao, and Yao]{Tong_ACS2018}
Tong,~Q.; Liu,~F.; Xiao,~J.; Yao,~W. Skyrmions in the moir{\'e} of van der {Waals} 2D Magnets. \emph{Nano Letters} \textbf{2018}, \emph{18}, 7194--7199\relax
\mciteBstWouldAddEndPuncttrue
\mciteSetBstMidEndSepPunct{\mcitedefaultmidpunct}
{\mcitedefaultendpunct}{\mcitedefaultseppunct}\relax
\EndOfBibitem
\bibitem[Hejazi \latin{et~al.}(2020)Hejazi, Luo, and Balents]{Hejazi_PNAS2020}
Hejazi,~K.; Luo,~Z.-X.; Balents,~L. Noncollinear phases in moir{\'e} magnets. \emph{Proceedings of the National Academy of Sciences} \textbf{2020}, \emph{117}, 10721--10726\relax
\mciteBstWouldAddEndPuncttrue
\mciteSetBstMidEndSepPunct{\mcitedefaultmidpunct}
{\mcitedefaultendpunct}{\mcitedefaultseppunct}\relax
\EndOfBibitem
\bibitem[Hejazi \latin{et~al.}(2021)Hejazi, Luo, and Balents]{Hejazi_PRB2021}
Hejazi,~K.; Luo,~Z.-X.; Balents,~L. Heterobilayer moir\'e magnets: Moir\'e skyrmions and commensurate-incommensurate transitions. \emph{Phys. Rev. B} \textbf{2021}, \emph{104}, L100406\relax
\mciteBstWouldAddEndPuncttrue
\mciteSetBstMidEndSepPunct{\mcitedefaultmidpunct}
{\mcitedefaultendpunct}{\mcitedefaultseppunct}\relax
\EndOfBibitem
\bibitem[Xiao \latin{et~al.}(2021)Xiao, Chen, and Tong]{Tong_PRR2021}
Xiao,~F.; Chen,~K.; Tong,~Q. Magnetization textures in twisted bilayer $\mathrm{Cr}\mathrm{X}_{3}$ ($\mathrm{X} = \mathrm{Br}, \mathrm{I})$. \emph{Phys. Rev. Research} \textbf{2021}, \emph{3}, 013027\relax
\mciteBstWouldAddEndPuncttrue
\mciteSetBstMidEndSepPunct{\mcitedefaultmidpunct}
{\mcitedefaultendpunct}{\mcitedefaultseppunct}\relax
\EndOfBibitem
\bibitem[Xiao \latin{et~al.}(2021)Xiao, Chen, and Tong]{Xiao_PRR2021}
Xiao,~F.; Chen,~K.; Tong,~Q. Magnetization textures in twisted bilayer $\mathrm{Cr}{\mathrm{X}}_{3}$ ($\mathrm{X}$=Br, I). \emph{Phys. Rev. Res.} \textbf{2021}, \emph{3}, 013027\relax
\mciteBstWouldAddEndPuncttrue
\mciteSetBstMidEndSepPunct{\mcitedefaultmidpunct}
{\mcitedefaultendpunct}{\mcitedefaultseppunct}\relax
\EndOfBibitem
\bibitem[Ghader \latin{et~al.}(2022)Ghader, Jabakhanji, and Stroppa]{Ghader_CommPhys2022}
Ghader,~D.; Jabakhanji,~B.; Stroppa,~A. Whirling interlayer fields as a source of stable topological order in moir{\'e} CrI3. \emph{Communications Physics} \textbf{2022}, \emph{5}, 192\relax
\mciteBstWouldAddEndPuncttrue
\mciteSetBstMidEndSepPunct{\mcitedefaultmidpunct}
{\mcitedefaultendpunct}{\mcitedefaultseppunct}\relax
\EndOfBibitem
\bibitem[Zheng(2023)]{Zheng_AFM2023}
Zheng,~F. Magnetic Skyrmion Lattices in a Novel 2D-Twisted Bilayer Magnet. \emph{Advanced Functional Materials} \textbf{2023}, \emph{33}, 2206923\relax
\mciteBstWouldAddEndPuncttrue
\mciteSetBstMidEndSepPunct{\mcitedefaultmidpunct}
{\mcitedefaultendpunct}{\mcitedefaultseppunct}\relax
\EndOfBibitem
\bibitem[Kim \latin{et~al.}(2023)Kim, Kiem, Bednik, Han, and Park]{Kim_NanoLett2023}
Kim,~K.-M.; Kiem,~D.~H.; Bednik,~G.; Han,~M.~J.; Park,~M.~J. Ab Initio Spin Hamiltonian and Topological Noncentrosymmetric Magnetism in Twisted Bilayer CrI3. \emph{Nano Letters} \textbf{2023}, \emph{23}, 6088--6094\relax
\mciteBstWouldAddEndPuncttrue
\mciteSetBstMidEndSepPunct{\mcitedefaultmidpunct}
{\mcitedefaultendpunct}{\mcitedefaultseppunct}\relax
\EndOfBibitem
\bibitem[Fumega and Lado(2023)Fumega, and Lado]{Fumega_2DMAtt2023}
Fumega,~A.~O.; Lado,~J.~L. Moiré-driven multiferroic order in twisted CrCl3, CrBr3 and CrI3 bilayers. \emph{2D Materials} \textbf{2023}, \emph{10}, 025026\relax
\mciteBstWouldAddEndPuncttrue
\mciteSetBstMidEndSepPunct{\mcitedefaultmidpunct}
{\mcitedefaultendpunct}{\mcitedefaultseppunct}\relax
\EndOfBibitem
\bibitem[Nica \latin{et~al.}(2023)Nica, Akram, Vijayvargia, Moessner, and Erten]{Nica_npjQM2023}
Nica,~E.; Akram,~M.; Vijayvargia,~A.; Moessner,~R.; Erten,~O. Kitaev spin-orbital bilayers and their moir\'e superlattices. \emph{npj Quantum Mater. 8, 9} \textbf{2023}, \relax
\mciteBstWouldAddEndPunctfalse
\mciteSetBstMidEndSepPunct{\mcitedefaultmidpunct}
{}{\mcitedefaultseppunct}\relax
\EndOfBibitem
\bibitem[Vijayvargia \latin{et~al.}(2023)Vijayvargia, Nica, Moessner, Lu, and Erten]{Vijayvargia_PRR2023}
Vijayvargia,~A.; Nica,~E.~M.; Moessner,~R.; Lu,~Y.-M.; Erten,~O. Magnetic fragmentation and fractionalized Goldstone modes in a bilayer quantum spin liquid. \emph{Phys. Rev. Res.} \textbf{2023}, \emph{5}, L022062\relax
\mciteBstWouldAddEndPuncttrue
\mciteSetBstMidEndSepPunct{\mcitedefaultmidpunct}
{\mcitedefaultendpunct}{\mcitedefaultseppunct}\relax
\EndOfBibitem
\bibitem[Xu \latin{et~al.}(2021)Xu, Ray, Shao, Jiang, Lee, Weber, Goldberger, Watanabe, Taniguchi, Muller, Mak, and Shan]{Xu_NatNano2021}
Xu,~Y.; Ray,~A.; Shao,~Y.-T.; Jiang,~S.; Lee,~K.; Weber,~D.; Goldberger,~J.~E.; Watanabe,~K.; Taniguchi,~T.; Muller,~D.~A.; Mak,~K.~F.; Shan,~J. Coexisting ferromagnetic--antiferromagnetic state in twisted bilayer CrI3. \emph{Nature Nanotechnology} \textbf{2021}, \relax
\mciteBstWouldAddEndPunctfalse
\mciteSetBstMidEndSepPunct{\mcitedefaultmidpunct}
{}{\mcitedefaultseppunct}\relax
\EndOfBibitem
\bibitem[Xie \latin{et~al.}(2023)Xie, Luo, Ye, Sun, Ye, Sung, Ge, Yan, Fu, Tian, Lei, Sun, Hovden, He, and Zhao]{Xie_NatPhys2022}
Xie,~H.; Luo,~X.; Ye,~Z.; Sun,~Z.; Ye,~G.; Sung,~S.~H.; Ge,~H.; Yan,~S.; Fu,~Y.; Tian,~S.; Lei,~H.; Sun,~K.; Hovden,~R.; He,~R.; Zhao,~L. Evidence of non-collinear spin texture in magnetic moir{\'e} superlattices. \emph{Nature Physics} \textbf{2023}, \relax
\mciteBstWouldAddEndPunctfalse
\mciteSetBstMidEndSepPunct{\mcitedefaultmidpunct}
{}{\mcitedefaultseppunct}\relax
\EndOfBibitem
\bibitem[Song \latin{et~al.}(2021)Song, Sun, Anderson, Wang, Qian, Taniguchi, Watanabe, McGuire, Stöhr, Xiao, Cao, Wrachtrup, and Xu]{Song_Science2021}
Song,~T.; Sun,~Q.-C.; Anderson,~E.; Wang,~C.; Qian,~J.; Taniguchi,~T.; Watanabe,~K.; McGuire,~M.~A.; Stöhr,~R.; Xiao,~D.; Cao,~T.; Wrachtrup,~J.; Xu,~X. Direct visualization of magnetic domains and moiré; magnetism in twisted 2D magnets. \emph{Science} \textbf{2021}, \emph{374}, 1140--1144\relax
\mciteBstWouldAddEndPuncttrue
\mciteSetBstMidEndSepPunct{\mcitedefaultmidpunct}
{\mcitedefaultendpunct}{\mcitedefaultseppunct}\relax
\EndOfBibitem
\end{mcitethebibliography}
\end{document}






\section{Details of the static spin structure factor calculation}

\begin{figure}[h]
\centering
   \includegraphics[width=1\textwidth]{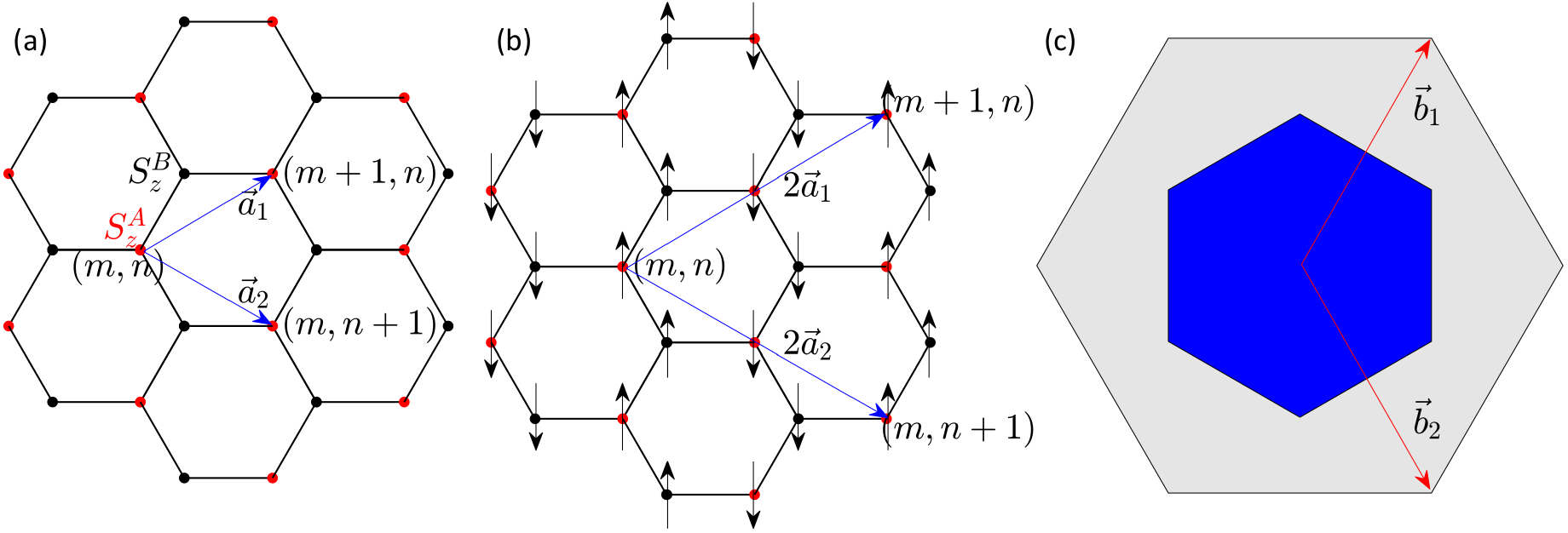}
    \caption{(a) Primitive vectors for the honeycomb lattice. (m,n) are integers and define the position of the site. If  $S_z^{A}=S_z^{B}=1$ and  $S_{x - y}^{A (B)}=0$ we have feromagnetic order. If $S_z^{A}=-S_z^{B}=1$ and  $S_{x - y}^{A (B)}=0$ we have antiferomagnetic order. (b) Ferromagnetic chain order in the zig-zag direction and primitive vectors defining unit cell in this ordering. (c) Reciprocal lattice with reciprocal lattice vectors. The blue region is the first Brillouin zone, }
    \label{Fig1}
\end{figure}
In this section, we derive form of the spin structure factor for different spin textures including FM, AFM and FMC. In all cases, we assume that the broken symmetry axis is along the $\hat{z}$ direction. We define the primitive lattice vectors for the honeycomb lattice as ${\textbf{a}}_1=\frac{3}{2}\hat{x}+\frac{\sqrt{3}}{2}\hat{y}$ and ${\textbf{a}}_2=\frac{3}{2}\hat{x}-\frac{\sqrt{3}}{2}\hat{y}$ shown in Fig. 1(a) (in units of bond length $a_0=1$). The corresponding reciprocal lattice vectors are ${\textbf{b}}_1=\frac{2\pi}{3}\hat{x}+\frac{2\pi}{\sqrt{3}}\hat{y}$ and ${\textbf{b}}_2=\frac{2\pi}{3}\hat{x}-\frac{2\pi}{\sqrt{3}}\hat{y}$ shown in Fig. 1(c).

We consider, $S_z^{A}=S_z^{B}=1$ for FM and $S_z^{A}=-S_z^{B}=1$ for AFM and calculate the spin structure factor as follows,
\begin{eqnarray}
S({\textbf{q}})&=&(1/N^2)\langle |\sum_{m,n}S_z^{A}e^{i\textbf{q}.\textbf{r}_{mn}}+S_z^{B}e^{i\textbf{q}.(\textbf{r}_{mn}+\frac{1}{2}\hat{x}+\frac{\sqrt{3}}{2}\hat{y})}|^2 \rangle \\ \nonumber
&=&\langle |\frac{1}{N}\sum_{m,n}e^{i\textbf{q}.\textbf{r}_{mn}}|^2\rangle \{(S_z^A)^2+(S_z^B)^2+2S_z^AS_z^B\cos(\frac{1}{2}q_x+\frac{\sqrt{3}}{2}q_y)\}  \\ \nonumber
&=&\delta_{{\textbf{q}},k_1{\bf{b}}_1+k_2{\bf{b}}_2}\{(S_z^A)^2+(S_z^B)^2+2S_z^AS_z^B\cos(\frac{1}{2}q_x+\frac{\sqrt{3}}{2}q_y)\}
\end{eqnarray}
where ${\bf{r}}_{mn}=m{\bf{a}}_1+n{\bf{a}}_2$ and $\{k_1, k_2, m,n  \} \in  \mathcal{Z}$. For FM ordering, the maximum value of $S({\bf{q}})$ is at the point ${\bf{q}}=(0,0)$, whereas for antiferromagnetic ordering, $S_z^{A}=-S_z^{B}=1$, the maximum value of $S({\bf{q}})$  is at the points ${\bf{q}}=\pm \{{\bf{b}}_1, {\bf{b}}_2, ({\bf{b}}_1+{\bf{b}}_2)\}$ in the region shown in Fig. 1(c).

For FMC, the lattice vectors defining the unit cell are $2{\bf{a}_1}$ and $2{\bf{a}_2}$ (see Fig. 1(b)). There are eight sites in the unit cell and the spin structure factor is
\begin{eqnarray}
S({\textbf{q}})&=&(1/N^2)\langle |\sum_{m,n}e^{i\textbf{q}.\textbf{r}_{mn}}+e^{i\textbf{q}.(\textbf{r}_{mn}+\frac{1}{2}\hat{x}+\frac{\sqrt{3}}{2}\hat{y})}+e^{i\textbf{q}.(\textbf{r}_{mn}+3\hat{x})}+e^{i\textbf{q}.(\textbf{r}_{mn}+\frac{7}{2}\hat{x}+\frac{\sqrt{3}}{2}\hat{y})}\\ \nonumber
&-&e^{i\textbf{q}.(\textbf{r}_{mn}+\frac{3}{2}\hat{x}+\frac{\sqrt{3}}{2}\hat{y})} -e^{i\textbf{q}.(\textbf{r}_{mn}+2\hat{x}+\sqrt{3}\hat{y})}-e^{i\textbf{q}.(\textbf{r}_{mn}+2\hat{x})}-e^{i\textbf{q}.(\textbf{r}_{mn}+\frac{3}{2}\hat{x}-\frac{\sqrt{3}}{2}\hat{y})}|^2\rangle \\ \nonumber
&=&\langle |\frac{1}{N}\sum_{m,n}e^{i\textbf{q}.\textbf{r}_{mn}}|^2\rangle\{16\sin^2(\frac{3}{4}q_x-\frac{\sqrt{3}}{4}q_y)(\sin(\frac{1}{2}q_x)+\sin(q_x+\frac{\sqrt{3}}{2}q_y))^2\} \\ \nonumber
&=&\delta_{{\textbf{q}},\frac{k_1}{2}{\bf{b}}_1+\frac{k_2}{2}{\bf{b}}_2}\{16\sin^2(\frac{3}{4}q_x-\frac{\sqrt{3}}{4}q_y)(\sin(\frac{1}{2}q_x)+\sin(q_x+\frac{\sqrt{3}}{2}q_y))^2\}
\end{eqnarray}
where  ${\bf{r}}_{mn}=2m{\bf{a}}_1+2n{\bf{a}}_2$ and and $\{k_1, k_2, m,n  \} \in  \mathcal{Z}$. The maximum value of $S({\bf{q}})$ occurs at the points ${\bf{q}}=\pm ({\bf{b}}_1+{\bf{b}}_2)/2$ in the region shown in Fig. 1(c) for the zigzag direction. Due to $C_3$ symmetry, for the ferromagnetic chain orders in the other directions the maximum values appears at the points ${\bf{q}}=\pm \{{\bf{b}}_1/2,{\bf{b}}_2/2\}$.